\documentclass[a4paper]{article}
\usepackage[utf8]{inputenc}
\usepackage{amsmath}
\usepackage{graphicx}
\usepackage{makeidx}  % allows for indexgeneration
\usepackage{longtable}
\usepackage{url}
\usepackage[margin=2cm]{geometry}

\graphicspath{{FIGS/}}

\usepackage{natbib}
\setcitestyle{authoryear,round,aysep={}}
\title{Stochastic Agent-Based Models of Intimate Partner Violence}
\author{Elisa Guidi$^{1,4}$, Patrizia Meringolo$^{2,4}$, Andrea Guazzini$^{2,4}$, Franco Bagnoli$^{3,4,5}$ \\[1cm]
\begin{minipage}{1.0\columnwidth}
\begin{enumerate}
\item Department of Information Engineering, University of Florence, Italy
\item Department of Education and Psychology, University of Florence, Italy
\item Department of Physics and Astronomy, University of Florence, Italy
\item INFN, sez. Firenze, Italy
\item Center for the Study of Complex Dynamics, University of Florence, Italy
\end{enumerate}
\end{minipage}
}

\begin{document}

\maketitle              % typeset the title of the contribution

\begin{abstract}
Intimate partner violence (IPV) is a significant public health problem and social issue that involves couples from all socioeconomic and cultural contexts. IPV may affect women and men, but these latter are the most common perpetrators of IPV. We developed stochastic Agent-Based models of IPV focused on the couple dynamics, determined by the parallel, individual behaviour of partners. Based on the psychological theory of the Cycle of Violence, we have developed a model based on four discrete states: passivity,  normal situation, upset and physical assault. The individual  transition probability depends on the previous state of the subject and that of the partner, and on a control parameter, the aggressiveness. We then let this parameter evolve depending on the perceived violence from past experiences (polarisation) or from the support received from the environment (social influence).  From the analysis of the phase diagrams we observe the emergence of characteristic patterns, in agreement with the observations of IPV in the literature.
\end{abstract}

%\keywords{Agent-Based Modeling; Violence; Psychology; Couple}
%
\section{Introduction}

The intimate relationships can be characterized by the presence of various forms of violence. In literature, intimate partner violence (IPV) refers to acts of physical, sexual, psychological violence and/or different types of controlling behaviours such as stalking inflicted by a current or previous intimate partner \citep{breiding2015intimate}.

Some studies showed that cohabiting couples are at an higher risk of IPV \citep{abramsky2011factors,Herrera2008}. A couple may be seen as a dynamic system in which partners’ interactions are influenced not only by the partner’s behaviours but also by contextual factors \citep{capaldi2007typological,katerndahl2010complex}. Indeed, a couple is not isolated but located in an environment (e.g., neighbours) and  may interact with different community members such as family members and friends, which have a critical role in maintaining or mitigating the issue of IPV \citep{goodman2011call,mancini2006preventing}.

Although the consequences of IPV are not symmetric between men and women,  the latter being more likely to suffer the most severe forms of violence and, therefore, of having more negative consequences (e.g., fear, injuries, and post-traumatic stress disorder) than men \citep{anderson2002perpetrator,caldwell2012gender}, it seems that there is not a profound difference between sexes regarding the factors leading to occasional acts of IPV \citep{marshall2011enduring}.  
Clearly, most of reports concerns the effects of repeated IPV, which corresponds almost invariably to men perpetrating violence against women.

\subsection{Factors related to  intimate partner violence: the role of social support}
Indeed, IPV is a complex and multi-dimensional problem and different researchers tried to describe its complexity \citep{ali2013intimate,KrugWHO2002}, highlighting a wide range of risk factors for IPV \citep{abramsky2011factors}

Among individual and relationship risk factors related to IPV, some researchers found that occurrence of men and women’s IPV was predicted by couple conflict, that is increased by individual variables such as hostility, depression and negative relationship attributions \citep{marshall2011enduring}. However, the frequency of IPV was predicted by hostility for men’s perpetration and by couple conflict for women’s perpetration, highlighting the different nature of IPV between men (i.e., individualized nature) and women (i.e., dyadic nature) \citep{marshall2011enduring}.

The occurrence of community violence and the presence of IPV in the social support networks are known community and societal risk factors of IPV, since in these cases people may have higher acceptance of violence (including IPV) and less possibility to receive a tangible support \citep{raghavan2006community,raiford2013interpersonal}. 

Indeed, social support has been recognized as a significant factor for IPV dynamics. As suggest by \citet{capaldi2012systematic}, social support has been investigated as a protective factor for victimization of female IPV. Victims of IPV are more likely to receive less social support than not abused ones \citep{katerndahl2013differences,van2003detangling}, and they are more likely to be re-abused if they have less social support \citep{bybee2005predicting,goodman2005women}. Additionally, some forms of social support are protective factors, reducing IPV perpetration \citep{slep2010unique}. However, \citet{katerndahl2013differences} hypothesized that social support may reduce the probability to be a victim of IPV but it may also allow a victim of IPV to remain in the violent relationship by decreasing the IPV impact on his/her mental health. 

Episodes of IPV can also happen in front of a third part defined in the literature as bystander \citep{hamby2016difference,planty2002third}, and IPV survivors usually report their experience of victimization to a member of their informal social support, who may react in a negative or unhealthy manner \citep{sylaska2014disclosure}. In the cases of IPV, among specific factors that may increase the likelihood to provide unhelpful support, anger seems to be associated with less helpful behaviours \citep{chabot2016beyond,chabot2009sex}.

One of the most important theoretical framework to understand the IPV is the ecological perspective \citep{ali2013intimate}, in which the causes of violence are not seen  as deterministic, but rather to occur probabilistically~\citep{heise2011works}. Applying Bronfenbrenner’s Ecological theory of human development \citep{bronfenbrenner1979ecology}, some researchers classified the IPV risk factors within four levels: individual, relationship, community and societal \citep{carlson1984causes,heise1998violence}. So far, among the risk factors that influence the rates of IPV, the majority of studies have more evaluate the individual and relationship levels rather than the community and societal levels \citep{KrugWHO2002}.

\subsection{Complexity science and dynamics patterns in intimate partner violence}

In the IPV literature, some theories have been focused on the dynamics of this issue such as the Cycle of Violence theory \citep{walker1979battered}, the Family Systems theory \citep{giles-sims1983systems}, and the Power and Control Wheel of the Duluth model \citep{pence1993education}. 

The Cycle of Violence theory \citep{walker1979battered} consists in three main phases that cyclically alternate: a) tension-building, where the abuser starts to be hostile with the partner, who tries to keep calm the aggressor; b) explosion, where the abuser starts to perpetrate violence; and c) honeymoon, during which the abuser starts to apologize for the behaviour and promises to stop to be violent \citep{ali2013intimate}. Within this violence cycle, there is a period called open window phase, between explosion and honeymoon stage, where it is more likely that the victim seeks help \citep{curnow1997open}.

In the Family Systems theory, after the first incident of IPV it is possible that the violence will stabilize or not, depending on positive or negative feedbacks received not only from the  internal system of the couple but also from the external context (e.g., family members, friends, etc.) \citep{giles-sims1983systems,katerndahl2014dynamics}.

In the Duluth model, the abuser uses different tactics to maintain the power and the control over the partner, and the sexual and physical violence may include all these tactics and IPV is not isolated incident but a constant presence for victims \citep{ali2013intimate,pence1993education}.

Recently, some researchers showed that these three theories of IPV dynamics may be expressed by different mathematical models of complex dynamics patterns \citep{burge2016using,katerndahl2014dynamics,katerndahl2012understanding,katerndahl2010complex}. In particular, these studies showed that the cycle of violence theory seems to be consisted with the periodic patterns (i.e., IPV is perceived as predictable); the family system theory yields   chaotic patterns (i.e., IPV is felt as less predictable than in the cycle of violence theory), and the Duluth model leads to the random patterns (i.e., IPV is  unpredictable) \citep{burge2016using,katerndahl2014dynamics,katerndahl2012understanding,katerndahl2010complex}. 

Actually, these three theories of IPV patterns are not mutually exclusive \citep{burge2016using}. Indeed, different patterns may be observed in different periods according with the availability of  resources, changes in the social connectedness or presence of  stress factors \citep{katerndahl2012understanding,katerndahl2014dynamics}.

\subsection{Agent-based modelling and couple dynamics}

Agent-based modelling (ABM) consists in modelling and simulating social systems by means of interacting agents that  follow simple probabilistic rules. The ABM approach has been used in different disciplines such as economics, social science and biology \citep{heath2009survey}, it should have more application in the field of social psychology \citep{smith2007agent,Bagnoli2007,Bagnoli2008,Bagnoli2014,Bagnoli2015,Bagnoli2016}. The ABM approach with its attention to micro and macro levels, non-linear effects, and multiple causal directions, is more able than prevalent approaches to describe emergent behaviour coming from interactive processes \citep{smith2007agent}.

In the literature, some studies have started to investigate, using the ABM,  couple processes such as the marital formation and dissolution \citep{mumcu2008marriage,saglam2013divorce}, and the seek-helping behaviours by women victims of IPV \citep{drigo2012modeling}. In particular, \citet{drigo2012modeling} highlighted that the ABM approach about IPV dynamics is suitable and furnishes implications for the policy.

\subsection{Aim of the study}

Following the suggestion by  \citet{smith2007agent}, we approach the dynamics of IPV by means of an agent-based model, in which the components of a couple can assume a finite number of states and each individual updates his/her state at discrete time steps in a probabilistic way, according with personal parameters, and based on the present state of the participants in the couple. 

We first define the transition probabilities and analyse the evolution of the couple in isolation. Since, as already said, the occasional presence of IPV seems to be influenced by the same factors for women and men \citep{marshall2011enduring}, we use a symmetric model for the two genders, although clearly they may be better represented by different values of the parameters. 

Secondly, we assume that the personal predisposition also evolves on the basis of  messages coming from the environment, assumed to be composed by similar couples (i.e., using a mean-field approximation). Beyond invoking uniformity, a justification of this approach is that couples tend to modify their network of contacts establishing links with other couples exhibiting similar behaviours such as women in violent relationship \citep{katerndahl2013differences}.

\section{Model 1: Short-time evolution after an upsetting episode}

Our first one aims at representing the short-time behaviour of a couple, starting from an upsetting episode and ending in an absorbing state like ``normal state'', predominant violence such as ``male violence'', ``female violence'', or ``mutual violence/separation''. Here we use the term ``predominant'' to indicate a situation in which the violence is mainly perpetuated by a single partner, while we use the term ``mutual'' when the violence is perpetuated by both members of the couple. Given that mutual violence may cause more injuries than non-reciprocal violence \citep{whitaker2007differences} and following past studies about help-seeking behaviour in the case of IPV \citep{ansara2010formal,douglas2012men}, we supposed that the victim experience mutual violence recognized the situation as severe and they will seek help and leave the relationship (e.g., separation). 

\begin{figure}
\begin{center}
\includegraphics[width=0.8\columnwidth]{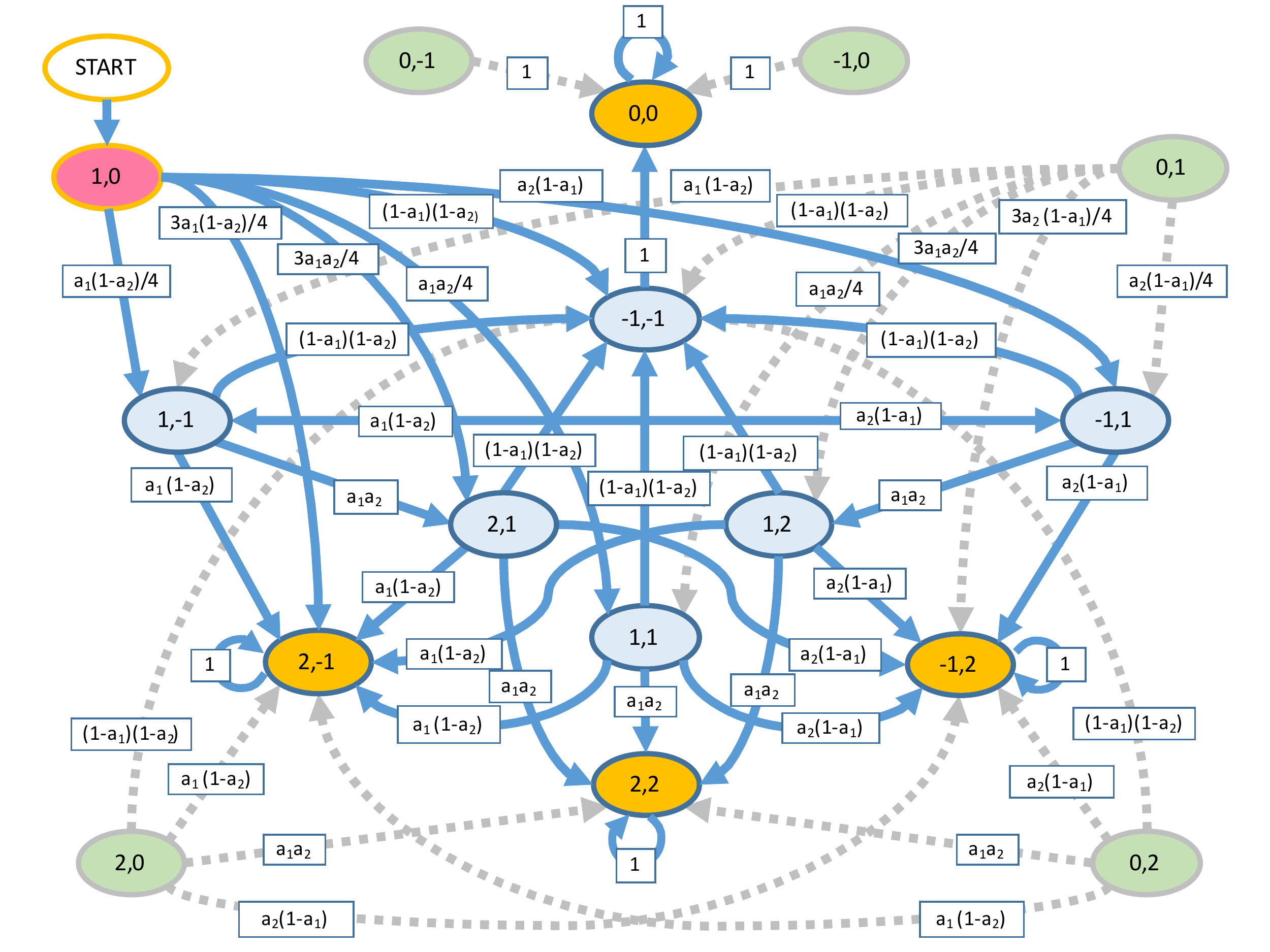} 
\end{center}
\caption [The transition diagram of couple dynamics for Model 1]{ The transition diagram of couple dynamics for Model 1. The ovals represent the 16 possible states $(s_1, s_2)$ of the couple and the arrows the transitions $M(s'_1, s_2|s_1, s_2; a_1, a_2)=\tau(s'1|s_1,s_2;a_1)\tau(s'_2|s_2, s_1;a_2)$. The initial state is coloured in red and marked by the START label. The green ovals are unreachable ``garden of Eden'' states, which can only be the starting states of the dynamics, and the corresponding transition probabilities are dashed. The four absorbing states normal $(0,0)$, separation $(2,2)$, male violence  $(2,-1)$,  female violence  $(-1,2)$
are marked in yellow. }
\label{fig:model1}
\end{figure}

\subsection{Model description}

We model the couple as composed by a man (opponent $1$) and a woman (opponent $2$), distinguished only by the fact that the evolution of the couple starts with the first one upset and the second one in a normal state. Each individual $i=1,2$ can assume four discrete states $s_i^t$ at time $t$, with $s_i \in \{-1,0,1,2\}$. We define these states following the Cycle of Violence Theory \citep{walker1979battered} as follows. The state   $s_i=0$ corresponds to the normal situation, while  $s_i=-1$ corresponds to passivity,  representing a situation of dependence and acceptance, but we also use this label to represent the ``beg for pardon'' state after an aggression. The label $s_i=1$ represents a tension condition where the member of the couple is upset, and finally $s_i=2$ corresponds to the presence of episodes of  violence or  physical assault. 

The model proceeds by discrete time steps. In each time step, the two individuals forming the couple face the other member and change his/her state (from $s_1$ and $s_2$ to $s'_1$ and $s'_2$ )  with a probability $\tau(s'_i|s_i, s_j; a)$, where $i$ represent the individual being updated and $j$ the partner and the parameter $a$ is described below.

Clearly, given a certain situation, the sum of all possible transition probabilities is one, i.e.,
\[
\sum_{s'_i=-1}^2 \tau(s'_i|s_i, s_j;a) = 1,
\]
for each $s_i$ and $s_j$. 

The transition matrix $\tau$ depends on a parameter $a$ (aggressiveness or assertiveness) that in our approximation represents, in a schematic manner, both the predisposition toward an aggressive behaviour, namely the tendency to attack the partner, and the active and assertive capacity of responding to the demands of partners, including also the ability to leave the relationship. We  use the same form of the transition matrix for both the male and the female members, possibly computed with different values of $a$. 

We divided the state of couple as a (tensor) product of individual states for two reasons. First of all, because in this way the model is more apt of being validated using personal profiles and secondly because for a couple there are 16 possible states (all combinations of the four individual states), which gives a transition matrix (from old to new states) with $16\times 16=256$ entries. At the individual level, the transition probability  $\tau(s'_i|s_i, s_j; a)$ has only $4^3=64$ entries, most of which are set to zero, as reported in the Appendix in Tables~\ref{tab:tau1.-1}--\ref{tab:tau1.2}. 

The basic idea is the following: for a low level of the aggressiveness factor $a$, the individual tends to return to the normal state $0$ or to the passive state $-1$ after an aggression. For high level of $a$ the individual tends to respond to an aggression becoming upset or responding with violence to violence. 

The transition probabilities of the couple, from $(s_1, s_2)$ to  $(s'_1, s'_2)$ with individual parameters $a_1$ and $a_2$ is given by 
\[
	M(s_1, s_2|s'_1, s'_2; a_1, a_2) = \tau(s'_1|s_1, s_2; a_1)\tau(s'_2|s_2, s_1; a_2).
\]
We can visualize the nonzero transition probabilities as a graph, as reported in Fig.~\ref{fig:model1}. 
The model presents four possible absorbing states, i.e: $(0,0)$, normal state of the couple, $(2,-1)$ and $(-1,2)$ which correspond to a situation in which one partner is violent and the other passive (prevarication), and $(2,2)$ in which both partners are violent and which is generally the prelude for the breaking of the couple. 

A time step is composed by two elementary processes that occur in parallel, for the two members of the couple. Each step is given by 
\begin{equation} \label{micro}
s'_i = \begin{cases} 
		-1 &\text{with probability $\tau(-1|s_i, s_j)$,}\\
		0 & \text{with probability $\tau(0|s_i, s_j)$,}\\
		1 & \text{with probability $\tau(1|s_i, s_j)$,}\\
		2 & \text{otherwise.}\\
\end{cases}
\end{equation} 

In practice, the choice of the new state, for example $s'_1$, was given by a random number $r$ between zero and one, that was confronted in sequence with the probability of the four possible outcomes: 
\begin{equation} 
\begin{cases}
s'_1 = -1 & \text{if $r < \tau(-1 | s_1,s_2),$} \\ 
s'_1 = 0 & \text{if $ \tau(-1 | s_1,s_2) \le r < \tau(-1 | s_1,s_2) + \tau(0 | s_1,s_2)$,} \\ 
s'_1 = 1 & \text{if $\tau(-1 | s_1,s_2) +\tau(0 | s_1,s_2) \le r < \tau(-1 | s_1,s_2) +\tau(0 | s_1,s_2)+\tau(1 | s_1,s_2)$,} \\ 
s'_1 = 2 & \text{otherwise, i.e., if $ r\ge   \tau(-1 | s_1,s_2) +\tau(0 | s_1,s_2)+\tau(1 | s_1,s_2)$.}
\end{cases}
\end{equation}

We started all simulations from a situation in which one partner (the male) is upset ($s_i^0=1)$ and the other is calm ($s_j^0=0$).

\begin{figure}[t]
\begin{verbatim}
a1=0.3, a2=0.3   % individual parameters
t=0, s1=1, s2=0  % initial state: mule upset and female normal
t=1, s1=1 s2=-1  % male upset and female remissive
t=2, s1=2 s2=1   % male violent and female upset
t=3, s1=2 s2=-1  % absorbing state: male violent and female passive
t=4, s1=2 s2=-1
t=5, s1=2 s2=-1
\end{verbatim}
\caption{An example of a stochastic trajectory.}\label{fig:traj}
\end{figure}
An example of a trajectory, for a high values of $a_1$ and an intermediate value of $a_2$ is reported in Figure~\ref{fig:traj}.
This trajectory can be read in this way: the male experiences a small inconvenient and becomes upset, while the female is calm ($t=0$). The male, due to his high aggressiveness $a_1$ maintains his state, while the female tries to calm him assuming a passive state ($t=1$). Instead of calming the partner, this passivity leads the male to assume a violent behaviour. In the meanwhile (the dynamics is parallel) the female, due to her intermediate value of $s_2$ (assertiveness in this case) becomes upset ($t=2$). However, confronted with violence and having an intermediate value of aggressiveness/assertiveness $a_2$ the female comes back to the passive state $s_2=-1$, while the male persists in his violent behaviour $s_1=2$. This configuration constitutes and absorbing state for the model. 
In this case the final state can be defined as male prevarication.

Clearly, the repetition of the simulation with the same parameters can lead to a different evolution, since the dynamics is stochastic. Hence, we should average over various realizations. It is however possible to obtain the evolution equation for the probability distribution for the couple (Markov chain).

Let us denote by $P(s_1,s_2;t)$, the probability of finding the couple in states ($s_1$, $s_2$) at time $t$. $P(s_1,s_2;t)$ has $16$ components, linked by the normalization condition 
\[
\sum_{s_1=-1}^2\sum_{s_2=-1}^2 P(s_1, s_2; t) =1.
\]

The temporal evolution of $P$ is given by the Markov equation
\begin{equation} 
P(s'_1,s'_2; t + 1) = \sum_{s_1}\tau_1(s'_1 | s_1,s_2;a_1)\tau_1(s'_2 | s_2,s_2;a_2) P(s_1,s_2;t). 
\label{markov}
\end{equation}

\subsection{Simulation results}

\begin{figure}
\begin{center}
\includegraphics[width=0.7\columnwidth]{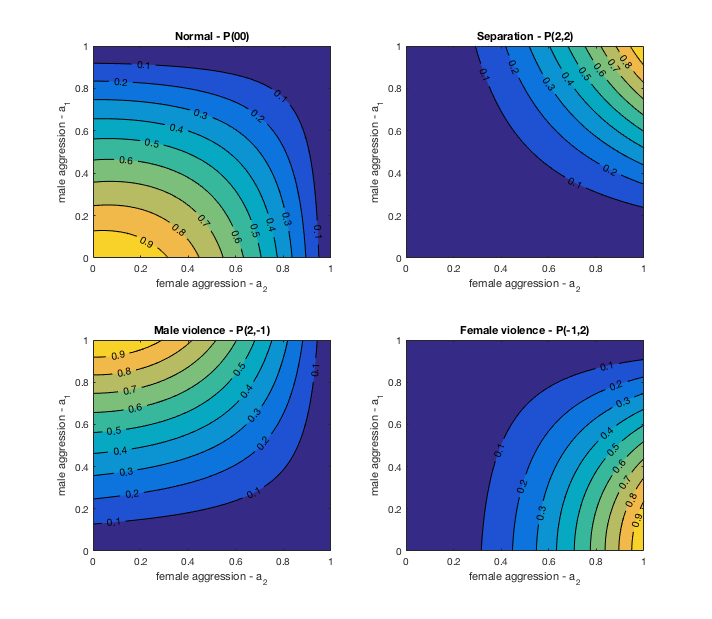} 
\end{center}
\caption [Basin of attraction of the four absorbing states of Model 1]{Basin of attraction of the four absorbing states of Model 1 for all possible values of male ($a_1$) and female ($a_2$) aggressiveness. The absorbing states are the asymptotic states of the probability distribution $P(s_1, s_2)$ corresponding to:  normal $P(0,0)$, separation  $P(2,2)$, male violence  $P(2,-1)$, female violence $P(-1,2)$. The asymmetry between male and female is only due to the initial state $P(1,0;0)=1$.}
\label{fig:model1-phase} 
\end{figure}

We repeated the simulation for all possible male and female aggressiveness, $a_1$ and $a_2$, we can obtain the phase diagram of the system, as reported in Fig.~\ref{fig:model1-phase}. In the Figure we report the probability of falling into an absorbing state (basin of attraction) starting with male upset ($P(1,0;0)=1$) for any value of the two aggressiveness parameters $a_1$ and $a_2$. In particular we show the asymptotic probability  $P(0,0)$ (normal behaviours),   $P(2,2)$ (mutual violence, leading to separation),   $P(2,-1)$ (male violence or prevarication),  $P(-1,2)$ (female violence).

The results are not unexpected. For low values of both male and female aggressiveness, the only asymptotic state is the ``quiescent'' one $(0,0)$. Similarly, for high values of both aggressiveness the only possible absorbing state is the mutual violence $(2,2)$, while with for two different values of the aggressiveness the final state is that of dominance.

\begin{figure}
\begin{center}
\includegraphics[width=0.5\columnwidth]{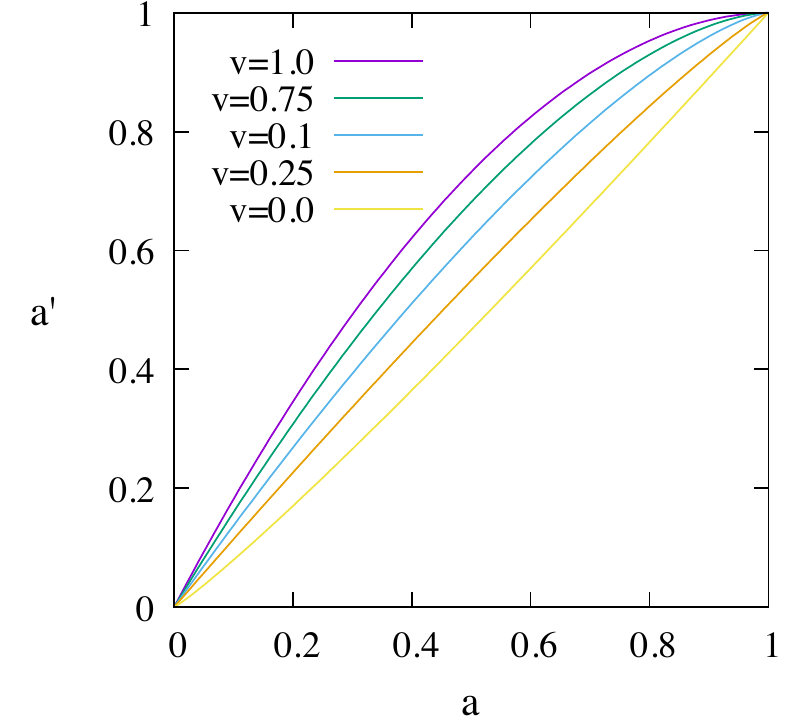} 
\end{center}
\caption [The evolution function of aggressiveness for different levels of perceived violence]{The evolution function of aggressiveness $a'=f(a;v,v_c)$ for different levels of perceived violence $v$, with $v_c=0.1$.}
\label{fig:ext1} 
\end{figure}

\begin{figure}
\begin{center}
\begin{tabular}{cc}
\includegraphics[width=0.48\columnwidth]{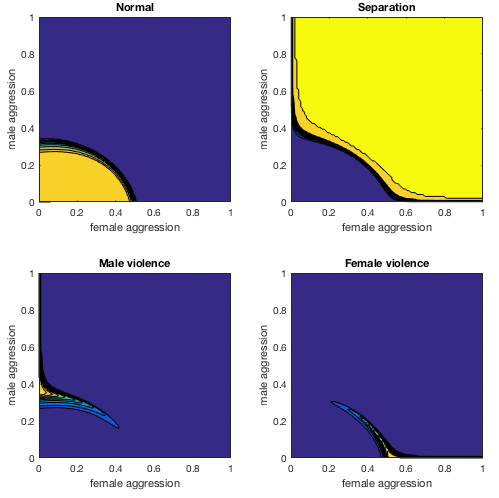} &
\includegraphics[width=0.49\columnwidth]{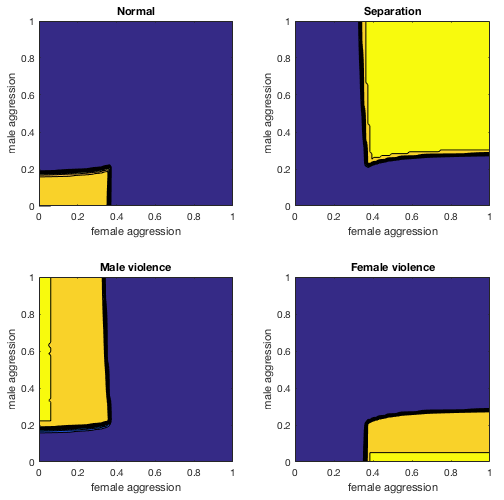} \\
no-gender-specific aggressiveness & gender-specific aggressiveness
\end{tabular}
\end{center}
\caption [Absorbing states of Model 1 with a mean-field evolution of the aggressiveness]{Absorbing states of Model 1 with a mean-field (self-consistent) evolution of the aggressiveness with $v_c=0.1$. Axes and plots as in Fig.~\ref{fig:model1-phase}. Averages over 20 runs.}
\label{fig:model2-f} 
\end{figure}

\section{Model 1 Self-consistent phase diagram}
Let us now explore the consequences of a social influence on the aggressiveness. 

\subsection{Model description}

We assume that a society is composed by a certain number of similar couples, all following the same dynamics. In other words, it is as if the couple were surrounded by ``mirrors'' reflecting their dynamics and influencing their own aggressiveness, i.e., a mean-field or self-consistent approach. 

We measure the perceived violence as the average number of violent states (2) assumed by one of the members of the couple after a certain number of time steps. In other words, we fix the parameters $a_1$ and $a_2$ and the initial state of the couple $P(1,0;0)=1$, let the system evolve for a number of time steps $T=20$ (generally sufficient to let the couple reach an absorbing state), after which we measure the gender violence $v_1$ and $v_2$ as 
\[
	\begin{split}
		v_1 &= P(2,-1;T)+P(2,2;T),\\
		v_2 &= P(-1,2;T)+P(2,2;T).
	\end{split}
\]

We then let both aggressiveness evolve depending on a threshold $v_c$: if the perceived violence is greater than the threshold the aggressiveness increases, the reverse in the opposite case
\begin{equation}
	a' = f(a; v, v_c) = 
	\begin{cases} 
	     1-(1-a)^{1+v-v_c} & \text{if $v>v_c$,}\\
     	a^{v_c-v+1} & \text{otherwise.}
   \end{cases}
   \label{f}
\end{equation}
The plot of the function $f(a; v, v_c)$ is reported in Fig.~\ref{fig:ext1} for $v_c=0.1$, value used in the simulations. The function is designed to provide a slow polarization of the aggressiveness (in both senses) according with the perceived violence in the environment.

\subsection{Simulation results}
The process is repeated $M=20$ times (turns). We studied two cases: one in which the perceived violence is not discriminated by gender, so that the value of the external perceived violence $v$  used in Eq.~\eqref{f} is simply the average of the two sexes $v=(v_1+v_2)/2$, and one in which the aggressiveness of each member of the couple evolves feeling only the appropriate gender violence. 

The results of simulations are reported in Figs.~\ref{fig:model2-f}. One can see that the situation is now much more extreme than in the simple case of Fig.~\ref{fig:model1-phase}, since no coexistence of phases is now possible: given an initial aggressiveness $a_1$ and $a_2$, the system almost always converges to a unique absorbing state. 

The other interesting aspect is the almost disappearance of the male and female prevarication if the perceived violence is ``asexual'', while the corresponding phases are much larger if the perceived violence only comes from the appropriate gender. This behaviour is sensible, albeit deviant: if male aggressiveness is only supported  by male violence, and  similarly for females,  ``cliques'' of similar behaviour can arise in the society.

\begin{figure}
\begin{center}
\includegraphics[width=0.7\columnwidth]{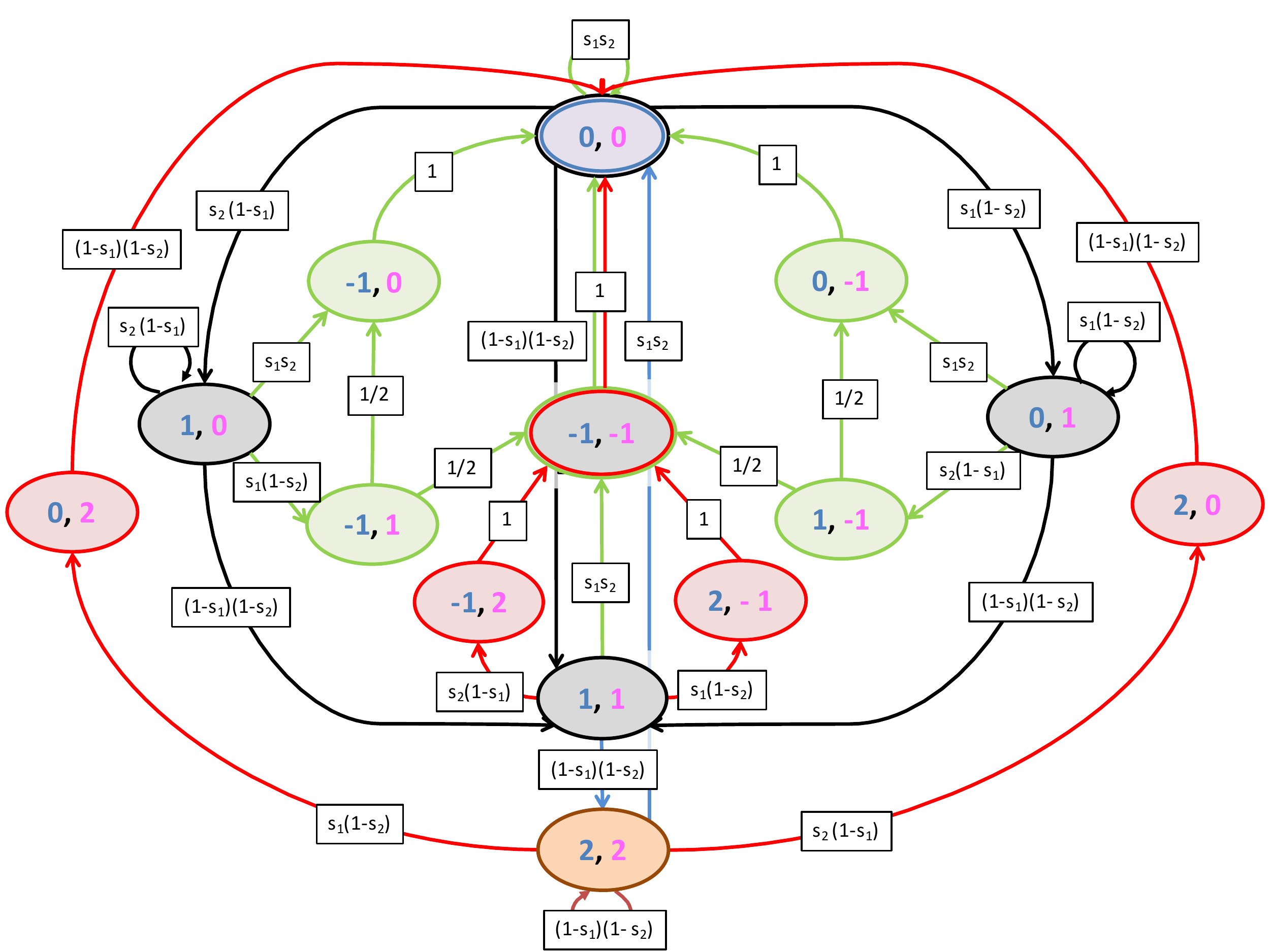} 
\end{center}
\caption [The transition matrix for Model 2]{The transition matrix for Model 2}
\label{graph}
\end{figure}

\begin{figure}
\begin{center}
\includegraphics[width=0.9\columnwidth]{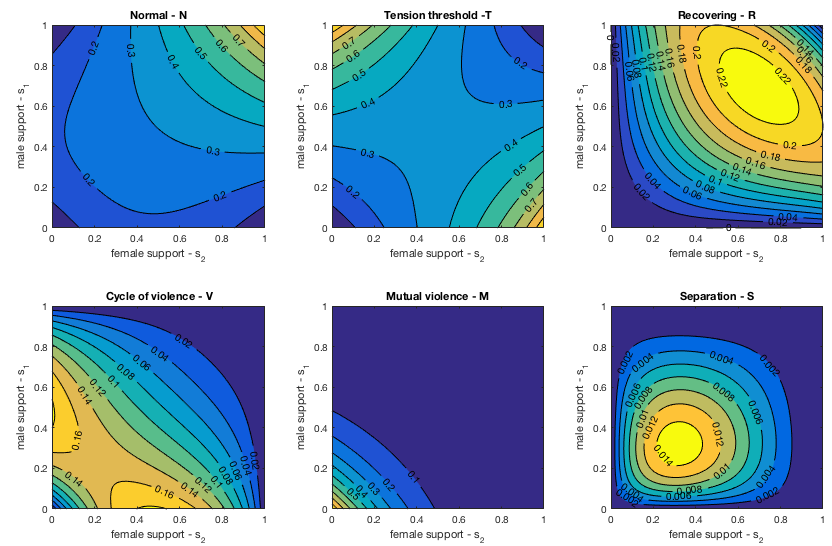} 
\end{center}
\caption [Probability of observing the normal behaviour, tension threshold, recovering path, cycle of violence, mutual violence and separation for Model 2]{Probability of observing the (from left to right and top to bottom) normal behaviour $\mathcal{N}$, tension threshold  $\mathcal{T}$, recovering path  $\mathcal{R}$, cycle of violence  $\mathcal{V}$, mutual violence  $\mathcal{M}$ and separation  $\mathcal{S}$ for a generic couple for Model 2}
\label{model2}
\end{figure}

\begin{figure}
\begin{center}
\includegraphics[width=0.5\columnwidth]{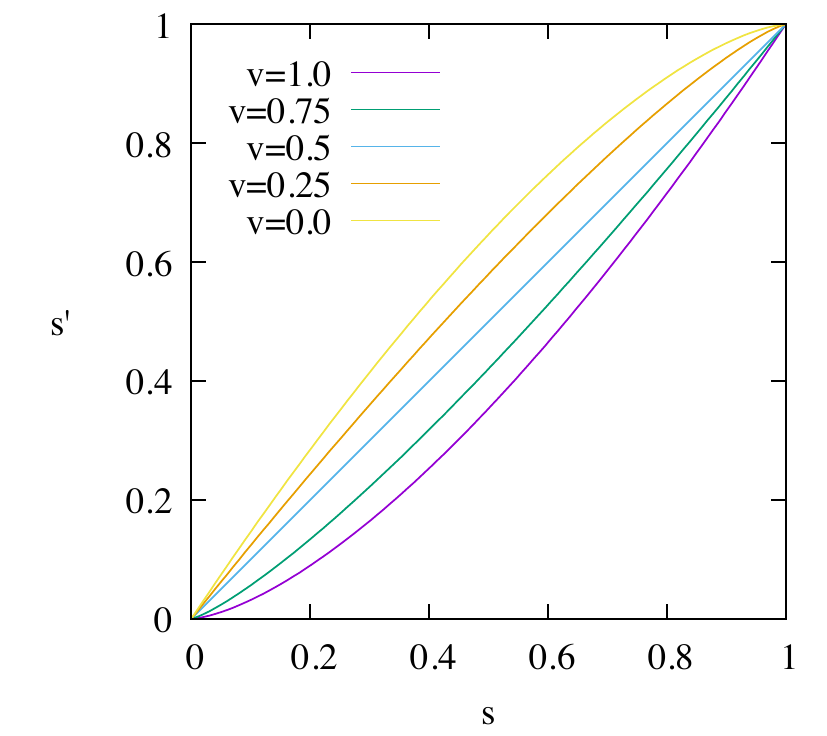} 
\end{center}
\caption [The evolution of support for Model 2 for different levels of perceived violence]{The evolution of support for Model 2  $s'=g(s;v, v_c)$ for different levels of perceived violence $v$ and $v_c=0.1$.}
\label{fig:ext}
\end{figure}

\section{Model 2: Long time-span behaviour of a typical couple}
The second model aims at representing the couple dynamics over a longer time span, so that for instance couples that reach the ``separation point'' of mutual violence are replaced with new couples initially in the calm state. This model is specifically aimed at studying the  effect of social support on the long-term dynamics.  

\subsection{Model description}

In this case we used as parameter the support received by the society, in the sense of reinforcement of assertiveness but also of aggressiveness. 
We modified the individual transition probabilities as shown in Tables~\ref{tab:tau3.-1}--\ref{tab:tau3.2}, reported in the Appendix. Given that anger seems to be associated with less helpful behaviours \citep{chabot2016beyond,chabot2009sex}, we suppose that parameter support $s$ is the opposite of parameter $a$. In this case we do not have any absorbing state.

The resulting transition graph for the evolution of the couple is shown in Fig.~\ref{graph}. We marked in red the paths that may lead to episodes of violence, in green those corresponding to normal behaviour with occasional upsetting episodes, and in grey the states belonging to both. Notice that we have two ``garden of Eden'' states, namely $(2,1)$ and $(1,2)$ that cannot be reached by dynamics and have been eliminated. 

We tried to measure the importance of the different paths, after a transient of $T=20$ steps (sufficient to reach an asymptotic state), starting from the ``male upset'' episode $P(1,0;0)=1$. With  ``normality'' we still refer to the asymptotic weight of state $\mathcal{N}=P(0,0)$. We then measured the ``threshold'' $\mathcal{T}$ condition (grey states in Fig.~\ref{graph}) as the weight of states $(1,0)$, $(0,1)$ plus the flux from state $(1,1)$ to $(0,0)$ though state $(2,2)$, i.e. 
\[
      \mathcal{T}=P(0,1)+P(1,0)+P(1,1).
\]

The ``recovering'' path $\mathcal{R}$ is marked in green in Fig.~\ref{graph} and computed as
\[
	\mathcal{R} = P(-1,0)+P(0,-1)+P(-1,1)+P(1,-1)+P(-1,-1)-\left[P(-1,2)+P(2,-1)\right].
\]

The ``violence'' cycle $\mathcal{V}$, in red in  Fig.~\ref{graph}, is defined as 
\[
	\mathcal{V} = P(-1,2)+P(2,-1)+P(0,2)+P(2,0).
\]

We also measured the ``mutual violence'' component $\mathcal{M}$ as
\[
  \mathcal{M}= P(2,2)(1-s1)(1-s2),
\]
and finally the separation rate $\mathcal{S}$ as
\[
   \mathcal{S}= P(2,2)s_1s_2
\]
  
\subsection{Simulation results}
The resulting phase diagram of the evolution of all possible supports received by males and females are reported in Fig.~\ref{model2}. As expected, the normal state corresponds to high support, while the tension state (border between normal and violence) corresponds to high support for a gender and low support for the other. 

Similarly, mutual violence occurs for low support for both sexes. The cycle of violence extends near the mutual violence zone, with asymmetric support while the recovering path  is near the normal state, with relatively high support for both genders. 

The separation (flux from violence to normal state) is somewhat complementary to the cycle of violence, and occurs for moderate support (hence the violence). The separation occurs when the two partners have similar support factor, i.e., it  is located near the diagonal of the phase diagram, while for the cycle of violence is favoured by asymmetric factors.

\begin{figure}
\begin{center}
\includegraphics[width=0.8\columnwidth]{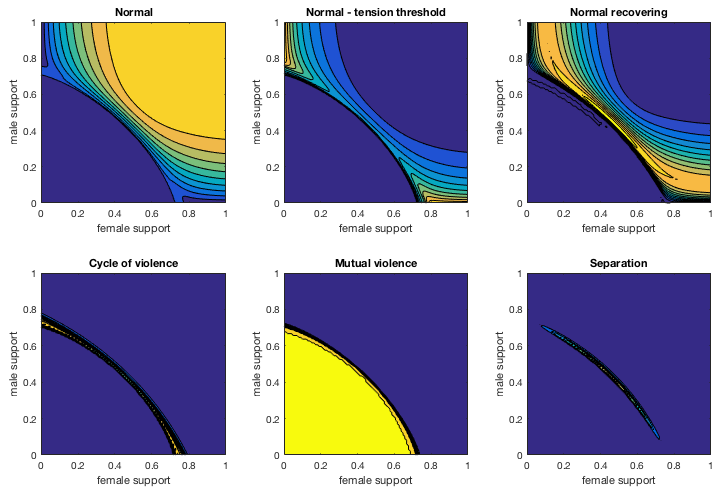} 
\end{center}
\caption [Phase diagram of Model 2 with a mean-field evolution of the aggressiveness: no gender perceived violence]{Phase diagram of Model 2 with a mean-field (self-consistent) evolution of the aggressiveness with $v_c=0.1$ and perceived violence not separated per gender. Axes and plots as in Fig.~\ref{model2}. Averages over 20 runs.}
\label{model2-nogender} 
\end{figure}

\begin{figure}
\begin{center}
\includegraphics[width=0.8\columnwidth]{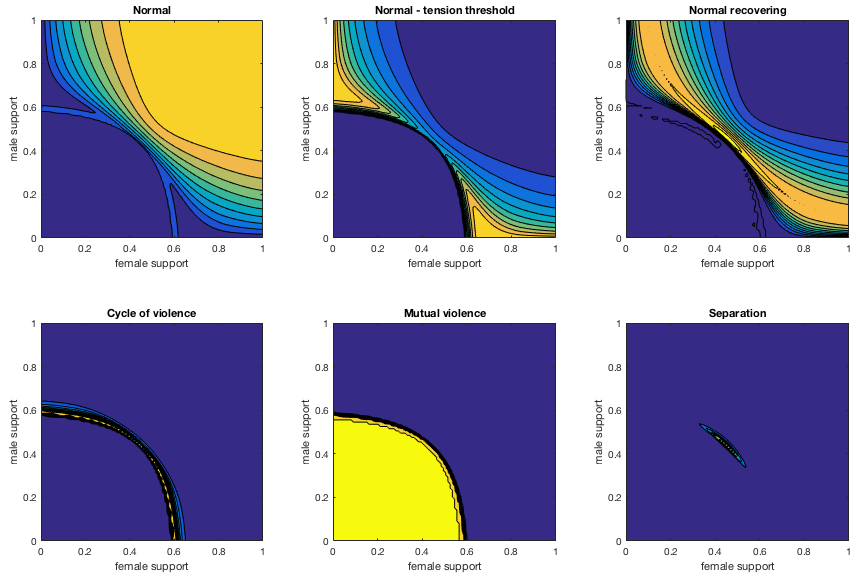} 
\end{center}
\caption [Phase diagram of Model 2 with a mean-field evolution of the aggressiveness: gender perceived violence]{Phase diagram of Model 2 with a mean-field (self-consistent) evolution of the aggressiveness with $v_c=0.1$ and gender perceived violence. Axes and plots as in Fig.~\ref{model2}. Averages over 20 runs.}
\label{model2-gender}
\end{figure}

\section{Model 2 Self-consistent phase diagram}
We apply here the same self-consistent approach as for Model 1, to our second model.

\subsection{Model description}

Given that presence of IPV in the social support networks may increase the acceptance of violence and decrease  the possibility to receive support \citep{raghavan2006community,raiford2013interpersonal}, we assume that the support $s$ evolves as a function of the perceived violence (see Fig.~\ref{fig:ext}) as

\begin{equation}
	s'=g(s; v, v_c) = 
		\begin{cases} 
	     s^{v-v_c+1} & \text{if $v>v_c$,}\\
     	1-(1-s)^{1+v_c-v} & \text{otherwise.}
   \end{cases}
   \label{g}
\end{equation}	

\subsection{Simulation results}
As in the previous case, the self-consistent behaviour is more polarized, even in the absence of absorbing states. The dominant states are now the mutual violence and the normal state, while the cycle of violence, and the separation phases are reduced. 

By comparing Figs~\ref{model2-nogender} and \ref{model2-gender}, it is evident that now the role of gender in the perceived violence (and thus in the evolution of the support) is marginal. 

\section{Conclusions}

In this paper, we described two stochastic agent-based models with the goal of investigating the dynamics of  intimate partner violence in a couple.

We first examined how the individual tendency to be aggressive and hostile towards the partner (i.e., the individual parameter), and the individual perception of violence in his/her social network (i.e., the contextual parameter) may give raise to intimate partner violence (i.e., emerging macroscopic social issue). 

Secondly, this ``short-time evolution'' model has been adapted to investigate the effect of an informal social support (i.e., a contextual parameter), developing a ``long-time span behaviour'' model.

Our first model foresaw the emergence of different absorbing states (e.g., `normal state'', ``male violence", ``female violence", or ``mutual violence/separation'') depending on the initial parameters (i.e.,  level of the aggressiveness factor $a$). Consistent with studies that highlighted how anger and hostility may increase the likelihood of perpetrating IPV \citep{norlander2005anger,shorey2011association}, simulation results of the first model showed  that high level of aggressiveness ($a$) in one member of the couple leads to a dominance pattern (e.g. , ``male violence'', ``female violence'') in which that individual is more likely to perpetrate violence than the other. Moreover, high level of aggressiveness in both sexes leads to the a reciprocal violence pattern  (e.g., ``mutual violence/separation''). 

Extending the first model by means of a polarization of the individual parameter ($a$) based on social influence of perceived violence in their context ($v$), simulation results showed an extremely clear distinction of couple behaviors. Interestingly, the male or female prevarication almost vanish if the perceived violence is ``asexual'', while the corresponding phases are much larger if the perceived violence only comes from the appropriate gender. A possible explanation of these results comes from social psychology which suggests that individuals follow social norms which define shared expectations about acceptable behavior in a society, proving individual behavior is regulated by social regulatory processes \citep{levine1990progress,sherif1967social,speltini1999gruppi}.

Despite the second model is more dynamic than the first, since it has not absorbent states, there are repeating patterns of behavior of the couple. Simulation results of this model showed that for a high symmetrical social support the couple has a higher likelihood to behave in a normal way with occasional conflicts that are resolved. In contrast with a low or an asymmetric social support,  violent patters, both in reciprocal and male or female violence, emerge more likely. As suggested by a recent review \citep{capaldi2012systematic}, the presence of social support may have a protective role for victimization and perpetration of IPV. Interestingly, after the occurrence of violence in the couple, if both members of the couple perceive a high social support, then the couple will have a recovering, while if they perceive a medium social support then the couple will leave. These results seem to support the hypothesis of \citet{katerndahl2013differences} that social support may decrease the chances to be a victim of IPV but it may also allow a victim of IPV to stay in the abuse relationship by reducing the IPV consequences (i.e., in this study the recovering condition). 

When we assumed that the support $s$ evolves as a function of the perceived violence, simulation results indicated a more polarized behaviors as the first model. However, contrary to the first model, the gender differences faded-out. These results emphasize that social support has a crucial role in preventing IPV, regardless of the sex of those who provide support to the woman or the man.

These findings also have practical implications. As suggested by \citet{Banyard2015} prevention interventions that based on giving community members a positive role of reducing IPV, such as bystander approach, should make individuals aware of being carriers of social norms related to IPV and they may modify them with their own behaviors in order to reduce violence in a society.
Given that some past studies showed that females are more likely to provide support in IPV situations \citep{banyard2008measurement,beeble2008factors}, it is important to engage more males in giving support towards individuals involved in IPV.

Future research should investigate the critical role of receiving social support after an episode of IPV given that it could increase the likelihood to remain in an abusive relationship. However, our study makes more evidences of the positive and protective role of social support within IPV dynamics.

Although in the literature there are few studies that have tried to investigate the dynamics of IPV through ABM, the models implemented in this study are a starting point for understanding the effect of social influence and social support on the dynamics of violence.

\section*{Acknowledgements}
F.B. acknowledges partial financial support from European
Commission (FP7-ICT-2013-10) Proposal No. 611299 SciCafe 2.0.

\bibliographystyle{jasss}
\bibliography{BIBLIO}

\begin{thebibliography}{57}
\providecommand{\natexlab}[1]{#1}
\providecommand{\url}[1]{\texttt{#1}}
\providecommand{\urlprefix}{URL }
\expandafter\ifx\csname urlstyle\endcsname\relax
  \providecommand{\doi}[1]{doi:\discretionary{}{}{}#1}\else
  \providecommand{\doi}{doi:\discretionary{}{}{}\begingroup
  \urlstyle{rm}\Url}\fi

\bibitem[{Abramsky et~al.(2011)Abramsky, Watts, Garcia-Moreno, Devries, Kiss,
  Ellsberg, Jansen \& Heise}]{abramsky2011factors}
Abramsky, T., Watts, C.~H., Garcia-Moreno, C., Devries, K., Kiss, L., Ellsberg,
  M., Jansen, H.~A. \& Heise, L. (2011).
\newblock What factors are associated with recent intimate partner violence?
  findings from the who multi-country study on women's health and domestic
  violence.
\newblock \textit{BMC public health}, \textit{11}(1), 1.
\newblock \doi{10.1186/1471-2458-11-109}
\newline\urlprefix\url{http://dx.doi.org/10.1186/1471-2458-11-109}

\bibitem[{Ali \& Naylor(2013)}]{ali2013intimate}
Ali, P.~A. \& Naylor, P.~B. (2013).
\newblock Intimate partner violence: A narrative review of the feminist, social
  and ecological explanations for its causation.
\newblock \textit{Aggression and Violent Behavior}, \textit{18}(6), 611--619.
\newblock \doi{10.1016/j.avb.2013.07.009}
\newline\urlprefix\url{http://dx.doi.org/10.1016/j.avb.2013.07.009}

\bibitem[{Anderson(2002)}]{anderson2002perpetrator}
Anderson, K.~L. (2002).
\newblock Perpetrator or victim? relationships between intimate partner
  violence and well-being.
\newblock \textit{Journal of Marriage and Family}, \textit{64}(4), 851--863.
\newblock \doi{10.1111/j.1741-3737.2002.00851.x}
\newline\urlprefix\url{http://dx.doi.org/10.1111/j.1741-3737.2002.00851.x}

\bibitem[{Ansara \& Hindin(2010)}]{ansara2010formal}
Ansara, D.~L. \& Hindin, M.~J. (2010).
\newblock Formal and informal help-seeking associated with women's and men's
  experiences of intimate partner violence in canada.
\newblock \textit{Social science \& medicine}, \textit{70}(7), 1011--1018.
\newblock \doi{10.1016/j.socscimed.2009.12.009}
\newline\urlprefix\url{http://dx.doi.org/10.1016/j.socscimed.2009.12.009}

\bibitem[{Bagnoli et~al.()Bagnoli, Carletti, Fanelli \& Guazzini}]{Bagnoli2007}
Bagnoli, F., Carletti, T., Fanelli, A., D.and~Guarino \& Guazzini, A. (????)

\bibitem[{Bagnoli et~al.(2008)Bagnoli, Guazzini \& Li{\`o}}]{Bagnoli2008}
Bagnoli, F., Guazzini, A. \& Li{\`o}, P. (2008).
\newblock Human heuristics for autonomous agents.
\newblock In Springer (Ed.), \textit{Bio-Inspired Computing and Communication
  LNCS 5151}, (p. 340).
\newblock \doi{10.1007/978-3-540-92191-2_30}
\newline\urlprefix\url{http://dx.doi.org/10.1007/978-3-540-92191-2_30}

\bibitem[{Banyard(2008)}]{banyard2008measurement}
Banyard, V.~L. (2008).
\newblock Measurement and correlates of prosocial bystander behavior: The case
  of interpersonal violence.
\newblock \textit{Violence and victims}, \textit{23}(1), 83--97.
\newblock \doi{doi:10.1891/0886-6708.23.1.83}
\newline\urlprefix\url{http://dx.doi.org/doi:10.1891/0886-6708.23.1.83}

\bibitem[{Banyard(2015)}]{Banyard2015}
Banyard, V.~L. (2015).
\newblock \textit{The Promise of a Bystander Approach to Violence Prevention},
  (pp. 7--23).
\newblock Cham: Springer International Publishing.
\newblock \doi{10.1007/978-3-319-23171-6_2}
\newline\urlprefix\url{http://dx.doi.org/10.1007/978-3-319-23171-6_2}

\bibitem[{Beeble et~al.(2008)Beeble, Post, Bybee \&
  Sullivan}]{beeble2008factors}
Beeble, M.~L., Post, L.~A., Bybee, D. \& Sullivan, C.~M. (2008).
\newblock Factors related to willingness to help survivors of intimate partner
  violence.
\newblock \textit{Journal of Interpersonal Violence}, \textit{23}(12),
  1713--1729.
\newblock \doi{10.1177/0886260508314333}
\newline\urlprefix\url{http://dx.doi.org/10.1177/0886260508314333}

\bibitem[{Breiding et~al.(2015)Breiding, Basile, Smith, Black \&
  Mahendra}]{breiding2015intimate}
Breiding, M., Basile, K., Smith, S., Black, M. \& Mahendra, R. (2015).
\newblock Intimate partner violence surveillance: Uniform definitions and
  recommended data elements, version 2.0.
\newblock Tech. rep., Atlanta, GA: National Center for Injury Prevention and
  Control, Centers for Disease Control and Prevention

\bibitem[{Bronfenbrenner(1979)}]{bronfenbrenner1979ecology}
Bronfenbrenner, U. (1979).
\newblock \textit{The ecology of human development: Experiments by nature and
  design}.
\newblock Cambridge: Harvard university press

\bibitem[{Burge et~al.(2016)Burge, Katerndahl, Wood, Becho, Ferrer \&
  Talamantes}]{burge2016using}
Burge, S.~K., Katerndahl, D.~A., Wood, R.~C., Becho, J., Ferrer, R.~L. \&
  Talamantes, M. (2016).
\newblock Using complexity science to examine three dynamic patterns of
  intimate partner violence.
\newblock \textit{Families, Systems, \& Health}, \textit{34}(1), 4--14.
\newblock \doi{10.1037/fsh0000170}
\newline\urlprefix\url{http://dx.doi.org/10.1037/fsh0000170}

\bibitem[{Bybee \& Sullivan(2005)}]{bybee2005predicting}
Bybee, D. \& Sullivan, C.~M. (2005).
\newblock Predicting re-victimization of battered women 3 years after exiting a
  shelter program.
\newblock \textit{American Journal of Community Psychology}, \textit{36}(1-2),
  85--96.
\newblock \doi{10.1007/s10464-005-6234-5}
\newline\urlprefix\url{http://dx.doi.org/10.1007/s10464-005-6234-5}

\bibitem[{Caldwell et~al.(2012)Caldwell, Swan \&
  Woodbrown}]{caldwell2012gender}
Caldwell, J.~E., Swan, S.~C. \& Woodbrown, V.~D. (2012).
\newblock Gender differences in intimate partner violence outcomes.
\newblock \textit{Psychology of Violence}, \textit{2}(1), 42--57.
\newblock \doi{10.1037/a0026296}
\newline\urlprefix\url{http://dx.doi.org/10.1037/a0026296}

\bibitem[{Capaldi \& Kim(2007)}]{capaldi2007typological}
Capaldi, D.~M. \& Kim, H.~K. (2007).
\newblock Typological approaches to violence in couples: A critique and
  alternative conceptual approach.
\newblock \textit{Clinical psychology review}, \textit{27}(3), 253--265.
\newblock \doi{10.1016/j.cpr.2006.09.001}
\newline\urlprefix\url{http://dx.doi.org/10.1016/j.cpr.2006.09.001}

\bibitem[{Capaldi et~al.(2012)Capaldi, Knoble, Shortt \&
  Kim}]{capaldi2012systematic}
Capaldi, D.~M., Knoble, N.~B., Shortt, J.~W. \& Kim, H.~K. (2012).
\newblock A systematic review of risk factors for intimate partner violence.
\newblock \textit{Partner abuse}, \textit{3}(2), 231--280.
\newblock \doi{10.1891/0886-6708.20.6.625}
\newline\urlprefix\url{http://dx.doi.org/10.1891/0886-6708.20.6.625}

\bibitem[{Carlson(1984)}]{carlson1984causes}
Carlson, B.~E. (1984).
\newblock Causes and maintenance of domestic violence: An ecological analysis.
\newblock \textit{The Social Service Review}, (pp. 569--587)
\newline\urlprefix\url{http://www.jstor.org/stable/30011762}

\bibitem[{Chabot et~al.(2016)Chabot, Gray, Makande \& Hoyt}]{chabot2016beyond}
Chabot, H.~F., Gray, M.~L., Makande, T.~B. \& Hoyt, R.~L. (2016).
\newblock Beyond sex likelihood and predictors of effective and ineffective
  intervention in intimate partner violence in bystanders perceiving an
  emergency.
\newblock \textit{Journal of interpersonal violence}, (p. 0886260515621064).
\newblock \doi{10.1177/0886260515621064}
\newline\urlprefix\url{http://dx.doi.org/10.1177/0886260515621064}

\bibitem[{Chabot et~al.(2009)Chabot, Tracy, Manning \& Poisson}]{chabot2009sex}
Chabot, H.~F., Tracy, T.~L., Manning, C.~A. \& Poisson, C.~A. (2009).
\newblock Sex, attribution, and severity influence intervention decisions of
  informal helpers in domestic violence.
\newblock \textit{Journal of Interpersonal Violence}, \textit{24}(10),
  1696--1713.
\newblock \doi{10.1177/0886260509331514}
\newline\urlprefix\url{http://dx.doi.org/10.1177/0886260509331514}

\bibitem[{Curnow(1997)}]{curnow1997open}
Curnow, S. A.~M. (1997).
\newblock The open window phase: Helpseeking and reality behaviors by battered
  women.
\newblock \textit{Applied Nursing Research}, \textit{10}(3), 128--135.
\newblock \doi{10.1016/S0897-1897(97)80215-7}
\newline\urlprefix\url{http://dx.doi.org/10.1016/S0897-1897(97)80215-7}

\bibitem[{Douglas et~al.(2012)Douglas, Hines \& McCarthy}]{douglas2012men}
Douglas, E.~M., Hines, D.~A. \& McCarthy, S.~C. (2012).
\newblock Men who sustain female-to-male partner violence: Factors associated
  with where they seek help and how they rate those resources.
\newblock \textit{Violence and victims}, \textit{27}(6), 871--894.
\newblock \doi{10.1891/0886-6708.27.6.871}
\newline\urlprefix\url{http://dx.doi.org/10.1891/0886-6708.27.6.871}

\bibitem[{Drigo et~al.(2012)Drigo, Ehlschlaeger \& Sweet}]{drigo2012modeling}
Drigo, M., Ehlschlaeger, C.~R. \& Sweet, E.~L. (2012).
\newblock Modeling intimate partner violence and support systems.
\newblock In D.~J. Westervelt \& L.~G. Cohen (Eds.),
  \textit{Ecologist-Developed Spatially-Explicit Dynamic Landscape Models},
  (pp. 235--253). Boston, MA: Springer US.
\newblock \doi{10.1007/978-1-4614-1257-1{\_}14}
\newline\urlprefix\url{http://dx.doi.org/10.1007/978-1-4614-1257-1{\_}14}

\bibitem[{Giles-Sims(1983)}]{giles-sims1983systems}
Giles-Sims, J. (1983).
\newblock \textit{Wife Battering: A Systems Theory Approach}.
\newblock New York, NY: Guilford Press

\bibitem[{Goodman et~al.(2005)Goodman, Dutton, Vankos \&
  Weinfurt}]{goodman2005women}
Goodman, L., Dutton, M.~A., Vankos, N. \& Weinfurt, K. (2005).
\newblock Women’s resources and use of strategies as risk and protective
  factors for reabuse over time.
\newblock \textit{Violence Against Women}, \textit{11}(3), 311--336.
\newblock \doi{10.1177/1077801204273297}
\newline\urlprefix\url{http://dx.doi.org/10.1177/1077801204273297}

\bibitem[{Goodman \& Smyth(2011)}]{goodman2011call}
Goodman, L.~A. \& Smyth, K.~F. (2011).
\newblock A call for a social network-oriented approach to services for
  survivors of intimate partner violence.
\newblock \textit{Psychology of Violence}, \textit{1}(2), 79--92.
\newblock \doi{10.1037/a0022977}
\newline\urlprefix\url{http://dx.doi.org/10.1037/a0022977}

\bibitem[{Guazzini et~al.(2015)Guazzini, Cini \& Bagnoli}]{Bagnoli2015}
Guazzini, A., Cini, A. \& Bagnoli, J., F.and~Ramasco (2015).
\newblock Opinion dynamics within a virtual small group: the stubbornness
  effect.
\newblock \textit{Frontiers in physics}, \textit{3}, 65.
\newblock \doi{10.3389/fphy.2015.00065}
\newline\urlprefix\url{http://dx.doi.org/10.3389/fphy.2015.00065}

\bibitem[{Hamby et~al.(2016)Hamby, Weber, Grych \&
  Banyard}]{hamby2016difference}
Hamby, S., Weber, M.~C., Grych, J. \& Banyard, V. (2016).
\newblock What difference do bystanders make? the association of bystander
  involvement with victim outcomes in a community sample.
\newblock \textit{Psychology of violence}, \textit{6}(1), 91--102.
\newblock \doi{10.1037/a0039073}
\newline\urlprefix\url{http://dx.doi.org/10.1037/a0039073}

\bibitem[{Heath et~al.(2009)Heath, Hill \& Ciarallo}]{heath2009survey}
Heath, B., Hill, R. \& Ciarallo, F. (2009).
\newblock A survey of agent-based modeling practices (january 1998 to july
  2008).
\newblock \textit{Journal of Artificial Societies and Social Simulation},
  \textit{12}(4), 9
\newline\urlprefix\url{http://jasss.soc.surrey.ac.uk/12/4/9.html}

\bibitem[{Heise(1998)}]{heise1998violence}
Heise, L.~L. (1998).
\newblock Violence against women an integrated, ecological framework.
\newblock \textit{Violence against women}, \textit{4}(3), 262--290.
\newblock \doi{10.1177/1077801298004003002}
\newline\urlprefix\url{http://dx.doi.org/10.1177/1077801298004003002}

\bibitem[{Heise(2011)}]{heise2011works}
Heise, L.~L. (2011).
\newblock \textit{What works to prevent partner violence? An evidence
  overview}.
\newblock London: STRIVE Research Consortium, London School of Hygiene and
  Tropical Medicine

\bibitem[{Herrera et~al.(2008)Herrera, Wiersma \& Cleveland}]{Herrera2008}
Herrera, V.~M., Wiersma, J.~D. \& Cleveland, H.~H. (2008).
\newblock The influence of individual and partner characteristics on the
  perpetration of intimate partner violence in young adult relationships.
\newblock \textit{Journal of Youth and Adolescence}, \textit{37}(3), 284--296.
\newblock \doi{10.1007/s10964-007-9249-4}
\newline\urlprefix\url{http://dx.doi.org/10.1007/s10964-007-9249-4}

\bibitem[{Katerndahl et~al.(2013)Katerndahl, Burge, Ferrer, Becho \&
  Wood}]{katerndahl2013differences}
Katerndahl, D., Burge, S., Ferrer, R., Becho, J. \& Wood, R. (2013).
\newblock Differences in social network structure and support among women in
  violent relationships.
\newblock \textit{Journal of Interpersonal Violence}, \textit{28}(9),
  1948--1964.
\newblock \doi{10.1177/0886260512469103}
\newline\urlprefix\url{http://dx.doi.org/10.1177/0886260512469103}

\bibitem[{Katerndahl et~al.(2014)Katerndahl, Burge, Ferrer, Becho \&
  Wood}]{katerndahl2014dynamics}
Katerndahl, D., Burge, S., Ferrer, R., Becho, J. \& Wood, R. (2014).
\newblock Dynamics of violence.
\newblock \textit{Journal of Evaluation in Clinical Practice}, \textit{20}(5),
  695--702.
\newblock \doi{10.1111/jep.12151}
\newline\urlprefix\url{http://dx.doi.org/10.1111/jep.12151}

\bibitem[{Katerndahl et~al.(2010)Katerndahl, Burge, Ferrer, Becho \&
  Wood}]{katerndahl2010complex}
Katerndahl, D.~A., Burge, S.~K., Ferrer, R.~L., Becho, J. \& Wood, R. (2010).
\newblock Complex dynamics in intimate partner violence: A time series study of
  16 women.
\newblock \textit{Primary care companion to the Journal of clinical
  psychiatry}, \textit{12}(4), e1--e12.
\newblock \doi{10.4088/PCC.09m00859whi}
\newline\urlprefix\url{http://dx.doi.org/10.4088/PCC.09m00859whi}

\bibitem[{Katerndahl et~al.(2012)Katerndahl, Burge, Ferrer, Becho \&
  Wood}]{katerndahl2012understanding}
Katerndahl, D.~A., Burge, S.~K., Ferrer, R.~L., Becho, J. \& Wood, R.~C.
  (2012).
\newblock Understanding intimate partner violence dynamics using mixed methods.
\newblock \textit{Families, Systems, \& Health}, \textit{30}(2), 141--153.
\newblock \doi{10.1037/a0028603}
\newline\urlprefix\url{http://dx.doi.org/10.1037/a0028603}

\bibitem[{Krug et~al.(2002)Krug, Dahlberg, Mercy, Zwi \& R}]{KrugWHO2002}
Krug, E., Dahlberg, L., Mercy, J., Zwi, A. \& R, L. (2002).
\newblock World report on violence and health.
\newblock Tech. rep., Geneva, World Health Organization

\bibitem[{Lauro~Grotto et~al.(2014)Lauro~Grotto, Guazzini \&
  Bagnoli}]{Bagnoli2014}
Lauro~Grotto, R., Guazzini, A. \& Bagnoli, F. (2014).
\newblock Metastable structures and size effects in small group dynamics.
\newblock \textit{Frontiers in psychology}, \textit{5}, 699.
\newblock \doi{10.3389/fpsyg.2014.00699}
\newline\urlprefix\url{http://dx.doi.org/10.3389/fpsyg.2014.00699}

\bibitem[{Levine \& Moreland(1990)}]{levine1990progress}
Levine, J.~M. \& Moreland, R.~L. (1990).
\newblock Progress in small group research.
\newblock \textit{Annual review of psychology}, \textit{41}(1), 585--634.
\newblock \doi{10.1146/annurev.ps.41.020190.003101}
\newline\urlprefix\url{http://dx.doi.org/10.1146/annurev.ps.41.020190.003101}

\bibitem[{Mancini et~al.(2006)Mancini, Nelson, Bowen \&
  Martin}]{mancini2006preventing}
Mancini, J.~A., Nelson, J.~P., Bowen, G.~L. \& Martin, J.~A. (2006).
\newblock Preventing intimate partner violence: A community capacity approach.
\newblock \textit{Journal of Aggression, Maltreatment \& Trauma},
  \textit{13}(3-4), 203--227.
\newblock \doi{10.1300/J146v13n03\_08}
\newline\urlprefix\url{http://dx.doi.org/10.1300/J146v13n03\_08}

\bibitem[{Marshall et~al.(2011)Marshall, Jones \&
  Feinberg}]{marshall2011enduring}
Marshall, A.~D., Jones, D.~E. \& Feinberg, M.~E. (2011).
\newblock Enduring vulnerabilities, relationship attributions, and couple
  conflict: an integrative model of the occurrence and frequency of intimate
  partner violence.
\newblock \textit{Journal of Family Psychology}, \textit{25}(5), 709--718.
\newblock \doi{10.1037/a0025279}
\newline\urlprefix\url{http://dx.doi.org/10.1037/a0025279}

\bibitem[{Mumcu \& Saglam(2008)}]{mumcu2008marriage}
Mumcu, A. \& Saglam, I. (2008).
\newblock Marriage formation/dissolution and marital distribution in a
  two-period economic model of matching with cooperative bargaining.
\newblock \textit{Journal of Artificial Societies and Social Simulation},
  \textit{11}(4), 3
\newline\urlprefix\url{http://jasss.soc.surrey.ac.uk/11/4/3.html}

\bibitem[{Norlander \& Eckhardt(2005)}]{norlander2005anger}
Norlander, B. \& Eckhardt, C. (2005).
\newblock Anger, hostility, and male perpetrators of intimate partner violence:
  A meta-analytic review.
\newblock \textit{Clinical psychology review}, \textit{25}(2), 119--152.
\newblock \doi{10.1016/j.cpr.2004.10.001}
\newline\urlprefix\url{http://dx.doi.org/10.1016/j.cpr.2004.10.001}

\bibitem[{Pence \& Paymar(1993)}]{pence1993education}
Pence, E. \& Paymar, M. (1993).
\newblock \textit{Education groups for men who batter: The Duluth model}.
\newblock New York, NY: Springer Publishing Company

\bibitem[{Planty(2002)}]{planty2002third}
Planty, M. (2002).
\newblock \textit{Third-party involvement in violent crime, 1993-99}.
\newblock Washington, DC: US Department of Justice, Office of Justice Programs,
  Bureau of Justice Statistics

\bibitem[{Raghavan et~al.(2006)Raghavan, Mennerich, Sexton \&
  James}]{raghavan2006community}
Raghavan, C., Mennerich, A., Sexton, E. \& James, S.~E. (2006).
\newblock Community violence and its direct, indirect, and mediating effects on
  intimate partner violence.
\newblock \textit{Violence Against Women}, \textit{12}(12), 1132--1149.
\newblock \doi{10.1177/1077801206294115}
\newline\urlprefix\url{http://dx.doi.org/10.1177/1077801206294115}

\bibitem[{Raiford et~al.(2013)Raiford, Seth, Braxton \&
  DiClemente}]{raiford2013interpersonal}
Raiford, J.~L., Seth, P., Braxton, N.~D. \& DiClemente, R.~J. (2013).
\newblock Interpersonal-and community-level predictors of intimate partner
  violence perpetration among african american men.
\newblock \textit{Journal of Urban Health}, \textit{90}(4), 784--795.
\newblock \doi{10.1007/s11524-012-9717-3}
\newline\urlprefix\url{http://dx.doi.org/10.1007/s11524-012-9717-3}

\bibitem[{Saglam(2013)}]{saglam2013divorce}
Saglam, I. (2013).
\newblock Divorce costs and marital dissolution in a one-to-one matching
  framework with nontransferable utilities.
\newblock \textit{Games}, \textit{4}(1), 106.
\newblock \doi{10.3390/g4010106}
\newline\urlprefix\url{http://dx.doi.org/10.3390/g4010106}

\bibitem[{Sherif(1967)}]{sherif1967social}
Sherif, M. (1967).
\newblock \textit{Social interaction: Process and products}.
\newblock Chicago: Aldine Publishing Company

\bibitem[{Shorey et~al.(2011)Shorey, Brasfield, Febres \&
  Stuart}]{shorey2011association}
Shorey, R.~C., Brasfield, H., Febres, J. \& Stuart, G.~L. (2011).
\newblock The association between impulsivity, trait anger, and the
  perpetration of intimate partner and general violence among women arrested
  for domestic violence.
\newblock \textit{Journal of Interpersonal Violence}, \textit{26}(13),
  2681--2697.
\newblock \doi{10.1177/0886260510388289}
\newline\urlprefix\url{http://dx.doi.org/10.1177/0886260510388289}

\bibitem[{Slep et~al.(2010)Slep, Foran, Heyman \& Snarr}]{slep2010unique}
Slep, A. M.~S., Foran, H.~M., Heyman, R.~E. \& Snarr, J.~D. (2010).
\newblock Unique risk and protective factors for partner aggression in a large
  scale air force survey.
\newblock \textit{Journal of Community Health}, \textit{35}(4), 375--383.
\newblock \doi{10.1007/s10900-010-9264-3}
\newline\urlprefix\url{http://dx.doi.org/10.1007/s10900-010-9264-3}

\bibitem[{Smith \& Conrey(2007)}]{smith2007agent}
Smith, E.~R. \& Conrey, F.~R. (2007).
\newblock Agent-based modeling: A new approach for theory building in social
  psychology.
\newblock \textit{Personality and Social Psychology Review}, \textit{11}(1),
  87--104.
\newblock \doi{10.1177/1088868306294789}
\newline\urlprefix\url{http://dx.doi.org/10.1177/1088868306294789}

\bibitem[{Speltini \& Palmonari(1999)}]{speltini1999gruppi}
Speltini, G. \& Palmonari, A. (1999).
\newblock \textit{I gruppi sociali}.
\newblock Bologna: Il mulino

\bibitem[{Sylaska \& Edwards(2014)}]{sylaska2014disclosure}
Sylaska, K.~M. \& Edwards, K.~M. (2014).
\newblock Disclosure of intimate partner violence to informal social support
  network members a review of the literature.
\newblock \textit{Trauma, Violence, \& Abuse}, \textit{15}(1), 3--21.
\newblock \doi{10.1177/1524838013496335}
\newline\urlprefix\url{http://dx.doi.org/10.1177/1524838013496335}

\bibitem[{Van~Wyk et~al.(2003)Van~Wyk, Benson, Fox \&
  DeMaris}]{van2003detangling}
Van~Wyk, J.~A., Benson, M.~L., Fox, G.~L. \& DeMaris, A. (2003).
\newblock Detangling individual-, partner-, and community-level correlates of
  partner violence.
\newblock \textit{Crime \& Delinquency}, \textit{49}(3), 412--438.
\newblock \doi{10.1177/0011128703049003004}
\newline\urlprefix\url{http://dx.doi.org/10.1177/0011128703049003004}

\bibitem[{Vilone et~al.(2016)Vilone, Carletti, Bagnoli \&
  Guazzini}]{Bagnoli2016}
Vilone, D., Carletti, T., Bagnoli, F. \& Guazzini, A. (2016).
\newblock The peace mediator effect: Heterogeneous agents can foster consensus
  in continuous opinion models.
\newblock \textit{J. Phys. A}, \textit{462}, 84.
\newblock \doi{10.1016/j.physa.2016.06.082}
\newline\urlprefix\url{http://dx.doi.org/10.1016/j.physa.2016.06.082}

\bibitem[{Walker(1979)}]{walker1979battered}
Walker, L.~E. (1979).
\newblock \textit{The battered woman}.
\newblock New York, NY: Harper and Row

\bibitem[{Whitaker et~al.(2007)Whitaker, Haileyesus, Swahn \&
  Saltzman}]{whitaker2007differences}
Whitaker, D.~J., Haileyesus, T., Swahn, M. \& Saltzman, L.~S. (2007).
\newblock Differences in frequency of violence and reported injury between
  relationships with reciprocal and nonreciprocal intimate partner violence.
\newblock \textit{American Journal of Public Health}, \textit{97}(5), 941--947.
\newblock \doi{10.2105/AJPH.2005.079020}
\newline\urlprefix\url{http://dx.doi.org/10.2105/AJPH.2005.079020}

\end{thebibliography}

\newpage
\section*{APPENDIX}

\begin{table}[h]
\caption [The transition matrix of Model 1 starting from state $-1$]{The transition matrix $\tau(s'_i|s_i, s_j)$ of Model 1 starting from state $s_i=-1$ (passive)}
\label{tab:tau1.-1} 
\begin{center}
\begin{tabular}{|c|c|c|c|l|}
\hline
$s'_i$ & $s_i$ & $s_j$ & $\tau(s'_i|s_i, s_j)$& $\qquad$ Illustration\\
%\caption{\label{tab:tau1} Tau1}\\
 \hline
 -1&-1&0 &0 & myself passive, spouse neutral $\rightarrow$ passive: not contemplated\\
 0&-1&0 &1 & myself passive, spouse neutral $\rightarrow$ normal: default\\
 1&-1&0 &0 & myself passive, spouse neutral $\rightarrow$ upset: not contemplated\\
 2&-1&0 &0 & myself passive, spouse neutral $\rightarrow$ violent: not contemplated\\
\hline
 -1&-1&-1 & 0 & myself passive, spouse passive $\rightarrow$ passive: not contemplated\\
 0&-1&-1 &1 & myself passive, spouse passive $\rightarrow$ normal: default\\
 1&-1&-1 &0 & myself passive, spouse passive $\rightarrow$ upset: not contemplated\\
 2&-1&-1 &0 & myself passive, spouse passive $\rightarrow$ violent: not contemplated\\
\hline
 -1&-1&1 &$1-a$ & myself passive, spouse upset $\rightarrow$ passive: not assertiveness\\
 0&-1&1 &0 & myself passive, spouse upset $\rightarrow$ normal: not contemplated\\
 1&-1&1 &$a$ & myself passive, spouse upset $\rightarrow$ upset: assertiveness\\
 2&-1&1 &0 & myself passive, spouse upset $\rightarrow$ violent: not contemplated\\
\hline
  %(-1,2) is an absorbing state implying the establishment of the real
  % violence cycle, although quite improbable
 -1&-1&2 &1 & myself passive, spouse violent $\rightarrow$ passive: passivity (absorbing state)\\
 0&-1&2 &0 & myself passive, spouse violent $\rightarrow$ normal: not contemplated\\
 1&-1&2 &0 &  myself passive, spouse violent $\rightarrow$ upset: not contemplated\\
 2&-1&2 &0 & myself passive, spouse violent $\rightarrow$ violent: not contemplated\\
 \hline
\end{tabular}
\end{center}
\end{table}

\begin{table}[h]
\caption[The transition matrix of Model 1 starting from state $0$]{The transition matrix $\tau(s'_i|s_i, s_j)$ of Model 1 starting from state $s_i=0$ (normal)}
\label{tab:tau1.0} 
\begin{center}
\begin{tabular}{|c|c|c|c|l|}
\hline
$s'_i$ & $s_i$ & $s_j$ & $\tau(s'_i|s_i, s_j)$& $\qquad$ Illustration\\
\hline
 % (0,0) i.e., (normal, normal) is an absorbing state
 -1&0&0 &0 & myself neutral, spouse neutral $\rightarrow$ passive: not contemplated\\
 0&0&0 &1 & myself neutral, spouse neutral $\rightarrow$ normal: normal (absorbing state)\\
  1&0&0& 0& myself neutral, spouse neutral $\rightarrow$ upset: not contemplated\\
  2&0&0 &0 & myself neutral, spouse neutral $\rightarrow$ violent: not contemplated\\
\hline
 -1&0&-1 &0 & myself neutral, spouse passive $\rightarrow$ passive: not contemplated\\
 0&0&-1 &1 & myself neutral, spouse passive $\rightarrow$ normal: default\\
 1&0&-1 &0 & myself neutral, spouse passive $\rightarrow$ upset: not contemplated\\
 2&0&-1 &0 & myself neutral, spouse passive $\rightarrow$ violent: not contemplated\\
\hline
 -1&0&1 &$1-a$ & myself neutral, spouse upset $\rightarrow$ passive: not assertiveness\\
 0&0&1 &0 & myself neutral, spouse upset $\rightarrow$ normal: not contemplated\\
 1&0&1 &a/4 & myself neutral, spouse upset $\rightarrow$ upset: assertiveness\\
 2&0&1 &3a/4 & myself neutral, spouse upset $\rightarrow$ violent: aggressiveness\\
\hline
 -1&0&2 &$1-a$ & myself neutral, spouse violent $\rightarrow$ passive: not aggressiveness\\
 0&0&2 &0 & myself neutral, spouse violent $\rightarrow$ normal: not contemplated\\
 1&0&2 &0 & myself neutral, spouse violent $\rightarrow$ upset: not contemplated\\
 2&0&2 &$a$ & myself neutral, spouse violent $\rightarrow$ violent: aggressiveness\\
\hline
\end{tabular}
\end{center}
\end{table}

\begin{table}[h]
\caption[The transition matrix of Model 1 starting from state $1$] {The transition matrix $\tau(s'_i|s_i, s_j)$ of Model 1 starting from state $s_i=1$ (upset)}
\label{tab:tau1.1} 
\begin{center}
\begin{tabular}{|c|c|c|c|l|}
\hline
$s'_i$ & $s_i$ & $s_j$ & $\tau(s'_i|s_i, s_j)$& $\qquad$ Illustration\\
%\caption{\label{tab:tau1} Tau1}\\
\hline
 -1&1&0 &$1-a$ & myself upset, spouse neutral $\rightarrow$ passive: not assertiveness\\
 0&1&0 &0 & myself upset, spouse neutral $\rightarrow$ normal: not contemplated\\
 1&1&0 &$a/4$ & myself upset, spouse neutral $\rightarrow$ upset: assertiveness\\
 2&1&0 &$3a/4$ & myself upset, spouse neutral $\rightarrow$ violent: aggressiveness\\
\hline
 -1&1&-1 &$1-a$ & myself upset, spouse passive $\rightarrow$ passive: not assertiveness\\
 0&1&-1 &0 & myself upset, spouse passive $\rightarrow$ normal: not contemplated\\
 1&1&-1 &0 & myself upset, spouse passive $\rightarrow$ upset: not contemplated\\
 2&1&-1 &a & myself upset, spouse passive $\rightarrow$ violent: aggressiveness\\
\hline
 -1&1&1 &$1-a$ & myself upset, spouse upset $\rightarrow$ passive: not assertiveness\\
 0&1&1 &0 & myself upset, spouse upset $\rightarrow$ normal: not contemplated\\
 1&1&1 &0 & myself upset, spouse upset $\rightarrow$ upset: not contemplated\\
 2&1&1 &$a$ & myself upset, spouse upset $\rightarrow$ violent: aggressiveness\\
\hline
 -1&1&2 &$1-a$ & myself upset, spouse violent $\rightarrow$ passive: passivity\\
 0&1&2 &0 & myself upset, spouse violent $\rightarrow$ normal: not contemplated\\
 1&1&2 &0 & myself upset, spouse violent $\rightarrow$ upset: default\\
 2&1&2 &$a$ & myself upset, spouse violent $\rightarrow$ violent: aggressiveness\\
\hline
\end{tabular}
\end{center}
\end{table}

\begin{table}[h]
\caption [The transition matrix of Model 1 starting from state $2$]{The transition matrix $\tau(s'_i|s_i, s_j)$ of Model 1 starting from state $s_i=2$ (violent)}
\label{tab:tau1.2} 
\begin{center}
\begin{tabular}{|c|c|c|c|l|}
\hline
$s'_i$ & $s_i$ & $s_j$ & $\tau(s'_i|s_i, s_j)$& $\qquad$ Illustration\\
%\caption{\label{tab:tau1} Tau1}\\
 \hline
 -1&2&0 &$1-a$ & myself violent, spouse neutral $\rightarrow$ passive: false passivity\\
 0&2&0 &0 & myself violent, spouse neutral $\rightarrow$ normal: not contemplated\\
 1&2&0 &0 & myself violent, spouse neutral $\rightarrow$ upset: not contemplated\\
 2&2&0 &$a$ & myself violent, spouse neutral $\rightarrow$ aggressiveness\\
 \hline

  % (2,-1) is an absorbing state implying the establishment of the real
  % violence cycle
 -1&2&-1&0 & myself violent, spouse passive $\rightarrow$ passive: not contemplated\\
 0&2&-1 &0 & myself violent, spouse passive $\rightarrow$ normal: non contemplated\\
 1&2&-1 &0 & myself violent, spouse passive $\rightarrow$ upset: not contemplated\\
 2&2&-1 &1 & myself violent, spouse passive $\rightarrow$ violent: aggressiveness (absorbing state)\\
\hline
 -1&2&1 &1-a & myself violent, spouse upset $\rightarrow$ passive: false passivity\\
 0&2&1 &0 & myself violent, spouse upset $\rightarrow$ normal: not contemplated\\
 1&2&1 &0 & myself violent, spouse upset $\rightarrow$ upset: not contemplated\\
 2&2&1 &a & myself violent, spouse upset $\rightarrow$ violent: aggressiveness\\
\hline
  % also (2,2) is an absorbing state, implying the end of the couple
 -1&2&2 &0 & myself violent, spouse violent $\rightarrow$ passive: not contemplated\\
 0&2&2 &0 & myself violent, spouse violent $\rightarrow$ normal: not contemplated\\
 1&2&2 &0 & myself violent, spouse violent $\rightarrow$ upset: not contemplated\\
 2&2&2 &1 & myself violent, spouse violent $\rightarrow$ violent: prelude to separation (absorbing state)\\
 \hline
\end{tabular}
\end{center}
\end{table}

\begin{table}[h]
\caption [The transition matrix of Model 2 starting from state $-1$]{The transition matrix $\tau_3(s'_i|s_i, s_j)$ of Model 2 starting from state $s_i=-1$ (passive)}
\label{tab:tau3.-1}
\begin{center}
\begin{tabular}{|c|c|c|c|l|}
\hline
$s'_i$ & $s_i$ & $s_j$ & $\tau_3(s'_i|s_i, s_j)$& $\qquad$ Illustration\\
%\caption{\label{tab:tau1} Tau1}\\
 \hline
  %%%%%%% stato iniziale: -1

  -1&-1&0 & 0 & myself passive, spouse neutral $\rightarrow$ passive: not contemplated\\
  0&-1&0 & 1 & myself passive, spouse neutral $\rightarrow$ neutral: default\\
  1&-1&0 & 0 & myself passive, spouse neutral $\rightarrow$ upset: not contemplated\\
  2&-1&0 & 0 & myself passive, spouse neutral $\rightarrow$ violent: not contemplated\\
 \hline
  -1&-1&-1 & 0 & myself passive, spouse passive $\rightarrow$ passive: not contemplated\\
  0&-1&-1 & 1 & myself passive, spouse passive $\rightarrow$ neutral: default\\
  1&-1&-1 & 0 & myself passive, spouse passive $\rightarrow$ upset: not contemplated\\
  2&-1&-1 & 0 & myself passive, spouse passive $\rightarrow$ violent: not contemplated\\
 \hline
  -1&-1&1 & 1 & myself passive, spouse upset $\rightarrow$ passive: default\\
  0&-1&1 & 0 & myself passive, spouse upset $\rightarrow$ neutral: not contemplated\\
  1&-1&1 & 0 & myself passive, spouse upset $\rightarrow$ upset: not contemplated\\
  2&-1&1 & 0 & myself passive, spouse upset $\rightarrow$ violent: not contemplated\\
 \hline
  % not contemplated, stato assorbente
  -1&-1&2 & 1 & myself passive, spouse violent $\rightarrow$ passive: default\\
  0&-1&2 & 0 & myself passive, spouse violent $\rightarrow$ neutral: not contemplated\\
  1&-1&2 & 0 & myself passive, spouse violent $\rightarrow$ upset: not contemplated\\
  2&-1&2 & 0 & myself passive, spouse violent $\rightarrow$ violent: not contemplated\\
 \hline
 \end{tabular}
 \end{center}
 \end{table}
 
 \begin{table}[h]
\caption [The transition matrix of Model 2 starting from state $0$]{The transition matrix $\tau_3(s'_i|s_i, s_j)$ of Model 2 starting from state $s_i=0$ (normal)}
\label{tab:tau3.0} 
\begin{center}
\begin{tabular}{|c|c|c|c|l|}
\hline
$s'_i$ & $s_i$ & $s_j$ & $\tau_3(s'_i|s_i, s_j)$& $\qquad$ Illustration\\
\hline

%%%%%%%%% stato iniziale: 0
  -1&0&0 & 0 & myself neutral, spouse neutral $\rightarrow$ passive: not contemplated\\
  0&0&0 & $s$ & myself neutral, spouse neutral $\rightarrow$ neutral: support\\
  1&0&0 & $1-s$ & myself neutral, spouse neutral $\rightarrow$ upset: negative episode + lack of support\\
  2&0&0 & 0 & myself neutral, spouse neutral $\rightarrow$ violent: not contemplated\\
 \hline
  -1&0&-1 & 0 & myself neutral, spouse passive $\rightarrow$ passive: not contemplated\\
  0&0&-1 & 1 & myself neutral, spouse passive $\rightarrow$ neutral: default\\
  1&0&-1 & 0 & myself neutral, spouse passive $\rightarrow$ upset: not contemplated\\
  2&0&-1 & 0 & myself neutral, spouse passive $\rightarrow$ violent: not contemplated\\
 \hline
  -1&0&1 & 0 & myself neutral, spouse upset $\rightarrow$ passive: not contemplated \\
  0&0&1 & $s$ & myself neutral, spouse upset $\rightarrow$ neutral: support\\
  1&0&1 & $1-s$ & myself neutral, spouse upset $\rightarrow$ upset: lack of support\\
  2&0&1 & 0 & myself neutral, spouse upset $\rightarrow$ violent: not contemplated\\
 \hline
  % 0,2 non si dovrebbe verificare mai: stato assorbente
  -1&0&2 & 0 & myself neutral, spouse violent $\rightarrow$ passive: not contemplated\\
  0&0&2 & 1 & myself neutral, spouse violent $\rightarrow$ neutral: default\\
  1&0&2 & 0 & myself neutral, spouse violent $\rightarrow$ upset: not contemplated\\
  2&0&2 & 0 & myself neutral, spouse violent $\rightarrow$ violent: not contemplated\\
 \hline
 \end{tabular}
 \end{center}
 \end{table}
 
 \begin{table}[h]
\caption [The transition matrix of Model 2 starting from state $1$]{The transition matrix $\tau_3(s'_i|s_i, s_j)$ of Model 2 starting from state $s_i=1$ (upset)}
\label{tab:tau3.1} 
\begin{center}
\begin{tabular}{|c|c|c|c|l|}
\hline
$s'_i$ & $s_i$ & $s_j$ & $\tau_3(s'_i|s_i, s_j)$& $\qquad$ Illustration\\
\hline
  %%%%%%% stato iniziale: 1

  -1&1&0 & $s$ & myself upset, spouse neutral $\rightarrow$ passive: support\\
  0&1&0 & 0 & myself upset, spouse neutral $\rightarrow$ neutral: not contemplated\\
  1&1&0 & $1-s$ & myself upset, spouse neutral $\rightarrow$ upset: lack of support\\
  2&1&0 & 0 & myself upset, spouse neutral $\rightarrow$ violent: not contemplated\\
 \hline
  -1&1&-1 & 1/2& myself upset, spouse passive $\rightarrow$ passive: may happen\\
  0&1&-1 & 1/2 & myself upset, spouse passive $\rightarrow$ neutral: may happen\\
  1&1&-1 & 0 & myself upset, spouse passive $\rightarrow$ upset: not contemplated\\
  2&1&-1 & 0 & myself upset, spouse passive $\rightarrow$ violent: not contemplated\\
 \hline
  -1&1&1 & $s$ & myself upset, spouse upset $\rightarrow$ passive: support\\
  0&1&1 & 0 & myself upset, spouse upset $\rightarrow$ neutral: not contemplated\\
  1&1&1 & 0 & myself upset, spouse upset $\rightarrow$ upset: not contemplated\\
  2&1&1 & $1-s$ & myself upset, spouse upset $\rightarrow$ violent: lack of support\\
 \hline
  % not contemplated, stato assorbente
  -1&1&2 & 0 & myself upset, spouse violent $\rightarrow$ passive: not contemplated\\
  0&1&2 & 0 & myself upset, spouse violent $\rightarrow$ neutral: not contemplated\\
  1&1&2 & 1& myself upset, spouse violent $\rightarrow$ upset: default\\
  2&1&2 & 0 & myself upset, spouse violent $\rightarrow$ violent: not contemplated\\
 \hline
 \end{tabular}
 \end{center}
 \end{table}
 
 \begin{table}[h]
\caption [The transition matrix of Model 2 starting from state $2$]{The transition matrix $\tau_3(s'_i|s_i, s_j)$ of Model 2 starting from state $s_i=2$ (violence)}
\label{tab:tau3.2} 
\begin{center}
\begin{tabular}{|c|c|c|c|l|}
\hline
$s'_i$ & $s_i$ & $s_j$ & $\tau_3(s'_i|s_i, s_j)$& $\qquad$ Illustration\\
\hline
  %%%%%%% stato iniziale: 2
  %uscita
  -1&2&0 & 0 & myself violent, spouse neutral $\rightarrow$ passive: not contemplated\\
  0&2&0 & 1 & myself violent, spouse neutral $\rightarrow$ neutral: default\\
  1&2&0 & 0 & myself violent, spouse neutral $\rightarrow$ upset: not contemplated\\
  2&2&0 & 0 & myself violent, spouse neutral $\rightarrow$ violent: not contemplated\\
 \hline
  %2,-1  diverso da femmina
  -1&2&-1 &1 & myself violent, spouse passive $\rightarrow$ passive: default\\
  0&2&-1 & 0 & myself violent, spouse passive $\rightarrow$ neutral: not contemplated\\
  1&2&-1 & 0 & myself violent, spouse passive $\rightarrow$ upset: not contemplated\\
  2&2&-1 & 0 & myself violent, spouse passive $\rightarrow$ violent: not contemplated\\
 \hline
  -1&2&1 & 0 & myself violent, spouse upset $\rightarrow$ passive: not contemplated\\
  0&2&1 & 0 & myself violent, spouse upset $\rightarrow$ neutral: not contemplated\\
  1&2&1 & 0 & myself violent, spouse upset $\rightarrow$ upset: not contemplated\\
  2&2&1 & 1 & myself violent, spouse upset $\rightarrow$ violent: default\\
 \hline
  % 2,2 diverso femmina
  -1&2&2 & 0 & myself violent, spouse violent $\rightarrow$ passive: not contemplated\\
  0&2&2 & $s$ & myself violent, spouse violent $\rightarrow$ neutral: prelude of separation (support)\\
  1&2&2 & 0 &  myself violent, spouse violent $\rightarrow$ upset: not contemplated\\
  2&2&2 & $1-s$ & myself violent, spouse violent $\rightarrow$ violent: increase of violence (lack of support)\\
 \hline
  \end{tabular}
 \end{center}
 \end{table}

\end{document}